\documentclass[fleqn,12pt]{wlscirep}
\usepackage{lineno}
\usepackage{setspace}
\usepackage{algorithm2e}

\title{Prudent behaviour accelerates disease transmission}

\author[1]{Samuel V. Scarpino}
\author[2]{Antoine Allard}
\author[1]{Laurent H\'{e}bert-Dufresne}
\affil[1]{Santa Fe Institute, Santa Fe, New Mexico, USA}
\affil[2]{Departament de F\'isica Fonamental, Universitat de Barcelona, Barcelona 08028, Spain}

\begin{document}

\maketitle
\thispagestyle{empty}
\doublespacing
\flushbottom
\textbf{Infectious diseases often spread faster near their peak than would be predicted given early data on transmission~\cite{}.  Despite the commonality of this phenomena, there are no known general mechanisms able to cause an exponentially spreading disease to begin spreading faster.  Indeed most features of real world social networks, e.g. clustering~\cite{volz2011effects, scarpino2015epidemiological} and community structure~\cite{salathe2010dynamics}, and of human behaviour, e.g. social distancing~\cite{glass2006targeted} and increased hygiene~\cite{fewtrell2005water}, will slow disease spread.  Here, we consider a model where individuals with essential societal roles--e.g. teachers, first responders, health-care workers, etc.-- who fall ill are replaced with healthy individuals. We refer to this process as \textit{relational exchange}.  Relational exchange is also a behavioural process, but one whose effect on disease transmission is less obvious.  By incorporating this behaviour into a dynamic network model, we demonstrate that replacing individuals can accelerate disease transmission.  Furthermore, we find that the effects of this process are trivial when considering a standard mass-action model, but dramatic when considering network structure.  This result highlights another critical shortcoming in mass-action models, namely their inability to account for behavioural processes.  Lastly, using empirical data, we find that this mechanism parsimoniously explains observed patterns across more than seventeen years of influenza and dengue virus data.  We anticipate that our findings will advance the emerging field of disease forecasting and will better inform public health decision making during outbreaks.}

Consider the school teacher who is infected with influenza by a student.  At some point, they may stay home from work due to the illness and a replacement instructor will fill their role.  For the ill teacher, this is likely a benefit, in that they have time to recover; however, the replacement teacher is now in a social situation where infection may be much more likely.  A similar narrative describes other essential members of society, such as health-care workers, police, firefighters, and public health officials.  Clearly such \textit{relational exchange} happens; nevertheless, it remains to be quantified how this behavioural process affects transmission and whether evidence exists for relational exchange in real world outbreaks.

In its most basic form, relational exchange is defined as a node replacement process where some individuals are deemed essential (e.g. teachers, janitors, health workers) and will be replaced by susceptible individuals if they are ever infected. This replacement process occurs at some rate, termed $\gamma$ in our equations, to account for a potential delay between when an individual becomes infectious and when they are diagnosed. Once replaced, the new susceptible individual is given some connections (e.g. students or patients) from the old essential individual. This relational exchange is important because: 1) the new susceptible node is introduced in to what is most likely a more dangerous situation with respect to disease risk and 2) bringing susceptible nodes from a different region of the contact network reduces the diameter of the population.

To begin, we investigated a standard mass-action model where nodes are distinguished by their state in a Susceptible-Infectious-Recovered (SIR) model. If we do not explicitly tag nodes as being essential or non-essential, we can effectively assume that each infectious node is potentially replaced by a susceptible node at rate $\gamma$. The epidemic is then described by system of equations~\ref{eq:massaction}. This model always leads to smaller epidemic peaks and final epidemic sizes than the equivalent model with no relational exchange (i.e., $\gamma = 0$). The logic behind this conclusion is straightforward, essentially the system behaves like a hybrid between SIR and Susceptible-Infectious-Susceptible (SIS) models with an effective recovery rate equal to the sum of the recovery and relational exchange ($\gamma$) rates.

What becomes apparent is that the critical feature of relational exchange, namely that the replacement individual is put into a more dangerous situation than they were in before, is simply not captured by mass-action models. Replacement individuals are thus not equivalent to a random susceptible individual. To properly account for this effect, we introduce network structure to the population (see Methods). In the network model, individuals in contact with infectious nodes will see their links rewired (the relation is exchanged) from an infectious to a susceptible node. Again, the events occur at a rate $\gamma$ for infectious nodes, and the dynamics otherwise follow the standard epidemiological model.  

\begin{figure}[t!!!!!!!]
\centering
\includegraphics[width=0.99\linewidth]{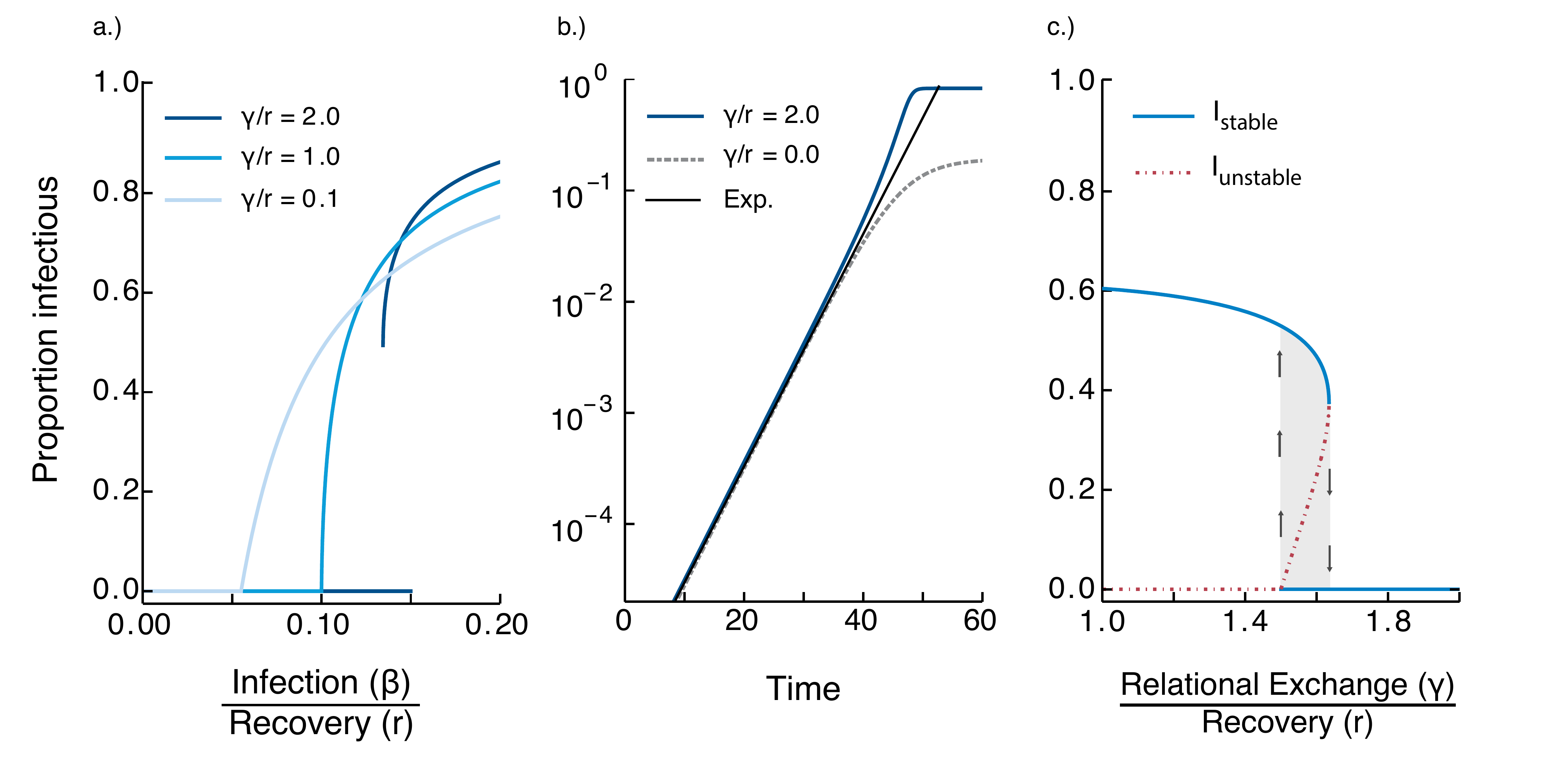}
\caption{\label{F1}\textbf{Analytical solutions of the relational exchange model on a network with average degree $\mathbf{\left\langle k\right\rangle = 20}$.} In panel \textit{a.)} we present the expected final epidemic size, which unlike the classic model, can undergo either a continuous or discontinuous transition at the epidemic threshold. In panel \textit{b.)} we present the time evolution at $\beta / r = 0.175$. The black line represents the expected exponential spread, whereas the dotted line shows the classic model and the continuous line the acceleration caused by relational exchange. Panel \textit{c.)} illustrates the hysteresis loop that can be encountered when varying the rates at which infectious individuals are replaced.}
\end{figure}

To facilitate analytic treatment, we consider the SIS model in system of equations~\ref{eq:unreduced_system}; however, qualitatively equivalent results apply to SIR models. Unlike with the mass-action model, here we find both that the final epidemic size and rate of spread just before the peak can be higher with relational exchange, Figure~\ref{F1}.  Put simply, relational exchange can parsimoniously account for faster transmission near an outbreak's peak than would be predicted given early data on transmission.  Importantly, the analytic results presented below represent the first mathematical investigation of such a behavioural model.

Furthermore, while the rules of the process are local, they model a global policy: the replacement rate is a function of how closely we are surveying the state of essential nodes, and of how quickly we wish to replace them. However, fast replacement rate is only effective at low prevalence, when secondary infections because of relational exchange are rare. Now, let us assume then that the replacement rate is strong enough to keep an outbreak under control, but that after some time the rate is slightly reduced.  This might occur, for example, after the initial fear wears off. Such a mechanism can push the system over a discontinuous transition, such that a \textit{microscopic} change in $\gamma$ can lead to a \textit{macroscopic} change in disease prevalence. We then wish to bring the system back to its previous state, so we increase $\gamma$ back to its previous value. Unfortunately, as shown in Figure~\ref{F1}c, the system exhibits a dangerous hysteresis loop. The replacement rate must be increased well beyond its previous value for the system to return to the initial state.

\begin{figure}[t!!!!!!!]
\centering
\includegraphics[width=0.99\linewidth]{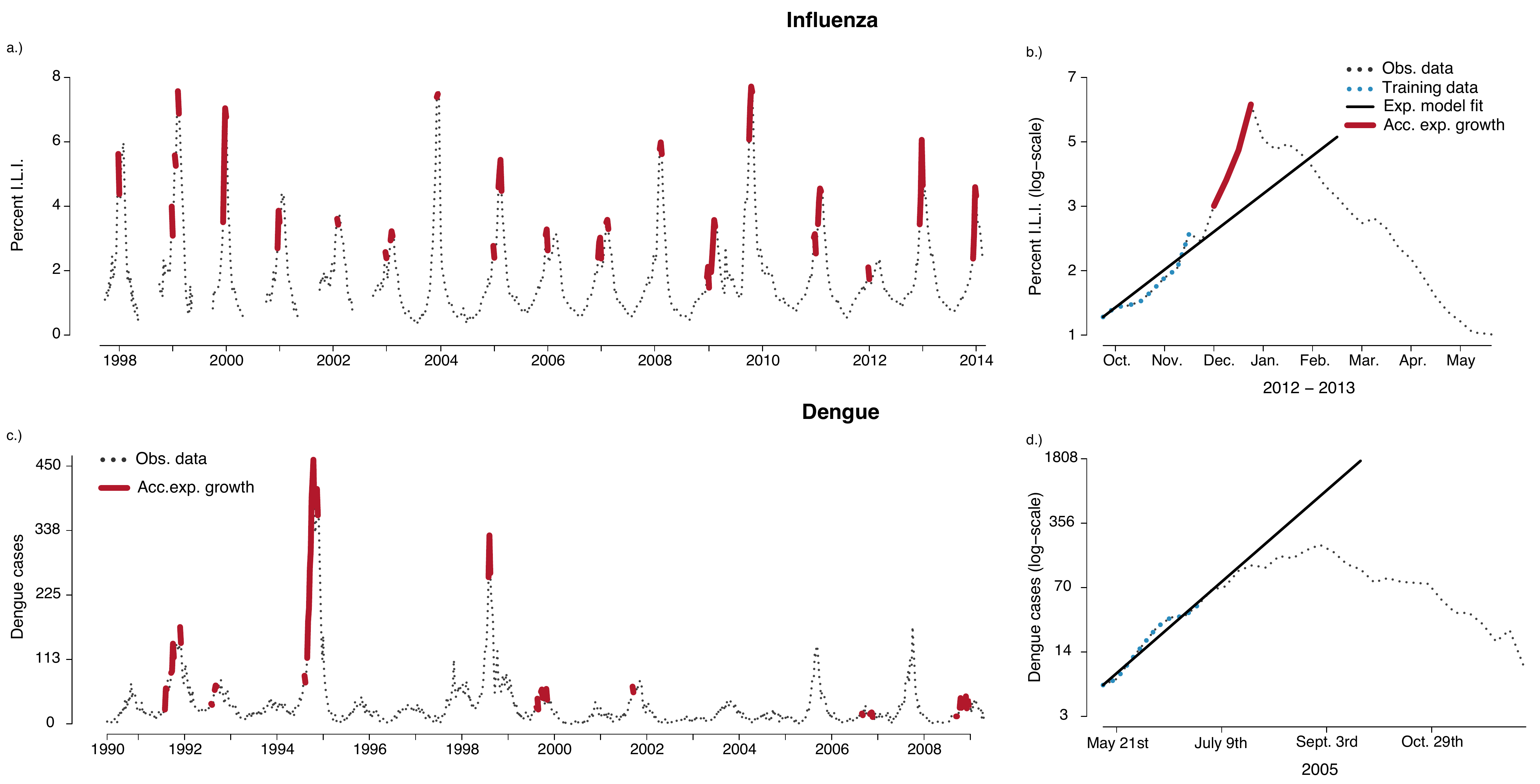}
\caption{\label{F2} \textbf{Empirical evidence for relational exchange.}  In panel \textit{a.)}, we examined seventeen years of national \textit{Influenza-Like-Illness} (I.L.I.) data in the U.S.A. (gray dashed lines) by fitting exponential growth models to the first eight weeks of the influenza season, defined by the U.S.A. Centers for Disease Control and Prevention as weeks 40 - 20, and then comparing the observed number of cases to the exponential prediction up until the peak.  For these data, I.L.I. was defined as a fever $\ge$ 100 degrees Fahrenheit (approx. 37.778 degrees Celsius) and one additional upper-respiratory symptom, e.g. sore throat, during the C.D.C. defined influenza season without another known non-influenza cause.  If more cases are reported than predicted by this exponential fit, we considered this evidence for accelerating spread and relational exchange (solid red lines).  We find evidence for relational exchange in all seventeen years and during the fall wave of the 2009 pandemic.  For dengue virus, panel \textit{c.)}, using fifteen years of dengue virus data reported to the U.S.A. C.D.C. in Puerto Rico, we performed the same analysis but found only a few isolated instances of accelerating spread.  Panels \textit{b.)} and \textit{d.)} show examples of the fitting process, with training data in blue and the exponential fit in solid black.}
\end{figure}

We now turn our attention to empirical data. The question we wish to answer is whether evidence exists for relational exchange in real-world diseases.  We select as our metric for relational exchange the presence of accelerating exponential transmission near the outbreak peak.  This phenomenon is a key feature of relational exchange and very uncommon in all other general models of disease spread published to date.  For this investigation, we selected two pathogens: influenza and dengue.  The rationale for selecting these diseases is that they both have strong seasonality--which can drive accelerating spread in some models\cite{colon, shaman}--and have rich historical data sets. However, because dengue is a vectored pathogen, and influenza is not, we expect the importance of relational exchange to be far greater for influenza~\cite{shaman2010absolute, johansson2009local, wearing2006ecological}.  To test for the presence of relational exchange, we fit exponential growth models to the first eight weeks of each season and then compare the predicted number of cases to the model fit up until the peak (see Methods).

In short, we find the predicted pattern: evidence for relational exchange for influenza, but not for dengue.  Figure~\ref{F2}, shows seventeen years of national-level influenza data from United States of America's Influenza-Like-Illness surveillance network.  In all seventeen years, and during the 2009 pandemic, we find strong evidence that the exponential rate of transmission accelerates before the peak.  For dengue, with over eighteen years of data, we see only a few idiosyncratic instances of accelerating transmission--aside from the 1995 outbreak, which was the most dramatic outbreak observed since 1990.  This observed difference between dengue and influenza can parsimoniously be accounted for by relational exchange.  To date, no alternative model can explain this difference, without substantially increased complexity or unrealistic assumptions.  Therefore, we conclude that evidence exists for relational exchange in real-world outbreaks.

This work is not without caveats.  First, if infectious individuals fail to transmit before being replaced, then relational exchange cannot increase epidemic size or transmission rate.  Second, in our analytical treatment, we assumed the population was infinitely large.  While this is a standard assumption, clearly the model cannot be directly applied to real-world populations.  Additionally, relaxing the infinite population size assumption is likely to exacerbate the negative effects of relational exchange by inducing small-worldness in the population~\cite{watts1998collective}.  Third, more complicated models, which included both seasonality and variable rates of infection between adults and children, could theoretically account for the empirical influenza observations.  However, such a model is unlikely to parsimoniously explain both influenza and dengue, as we have done with relational exchange.  Finally, we only considered two diseases in our empirical investigation.  Future work should consider classifying a variety of different diseases based on whether they exhibit evidence for relational exchange.

There are three additional implications of this study, which will affect the broader scientific, medical, and public health communities.  First, as recently demonstrated by Leventhal et al. (2015), the evolution of increased pathogen transmissibility decreases as a pathogen spreads through a heterogeneous population~\cite{leventhal2015evolution}.  This occurs because highly connected individuals are typically infected early in an outbreak and, as these individuals recover, the epidemic potential of the population decreases.  Therefore, any process, such as relational exchange, which maintains highly connected, susceptible individuals can increase the chance of more transmissible strains evolving and persisting.  Second, if individuals entering high-risk societal roles can be vaccinated or selected from resistant individuals, then relational exchange will no longer have negative effects on outbreak progression. Finally, methods for forecasting disease spread must include both realistic population structure and salient aspects of human behaviour.  Without these key features, we cannot hope for robust, actionable models for predicting epidemics.  

\section*{Methods}
\subsection*{Mass-action compartmental approach}
We use a standard mass-action model where nodes are distinguished by their state in a Susceptible-Infectious-Recovered dynamics (variables in brackets correspond to the fraction of the population in each state). The epidemic is described by the following system of equations,
\begin{subequations}\label{eq:massaction}
\begin{eqnarray}
\dot{[S]} & = & \gamma [I] - \beta [S][I] \\
\dot{[I]} & = & \beta [S][I] - \left(r + \gamma\right) [I] \\
\dot{[R]} & = & r[I] \; .
\end{eqnarray}
\end{subequations}
where $r$ is the recovery rate, $\beta$ is the transmission rate, and $\gamma$ is the replacement rate. All variables in bracket are dynamical, and we use the dot notation to identify time derivative.

\subsection*{Network model with pair approximation}
To adapt the relational exchange mechanism to network structures, we introduce pair approximations into the ODE system~\ref{eq:massaction}. Furthermore, and to allow for simpler analytical treatment, we consider SIS dynamics. We now write:

\begin{subequations} \label{eq:unreduced_system}
\begin{align}
  \dot{[S]} & = r [I] - \beta \left[ SI \right] \label{eq:Sdot} \\
  \dot{[I]} & = \beta \left[ SI \right] - r [I] \; , \label{eq:Idot} 
\end{align}
which is the standard SIS model, but written using \textit{pairs} $\left[ XY \right]$, defined as the \emph{per capita} number of links between nodes in state $X$ and $Y$. The time evolution of these pairs are ruled by
\begin{align}
  \dot{\left[SS\right]} & = \left(r+\gamma\right) \left[SI\right] - 2\beta \left[SI\right] \frac{\left[SS\right]}{[S]} \label{eq:SSdot} \\
  \dot{\left[SI\right]} & = \left(r+\gamma\right) \left(2\left[II\right] - \left[SI\right]\right) + \beta \left[SI\right] \left( 2\frac{\left[SS\right]}{[S]} - \frac{\left[SI\right]}{[S]} - 1\right)  \ , \label{eq:SIdot} \\
  \dot{\left[II\right]} & = \beta \left[SI\right] \left( 1 + \frac{\left[SI\right]}{[S]} \right) - 2 \left(r+\gamma\right) \left[II\right]. \label{eq:IIdot}
\end{align}
\end{subequations}

In this model, node states do not change following replacement (see absence of $\gamma$ in $\dot{[S]}$ and $\dot{[I]}$). However, links are rewired: e.g., $\left[SI\right]$ can become $\left[SS\right]$ if the infectious individual is replaced. The result is an adaptive network model where essential relations (e.g., patient-doctor, student-teacher) can be rewired.

This model is similar to a variant of the adaptive network model introduced by Gross et al. (2006) ~\cite{gross2006epidemic}. However, the adaptive process of Gross et al. (2006) \textit{always} hinders disease spread as the only possible adaptive move is $\left[SI\right]$ to $\left[SS\right]$. In our model, there is an interesting trade-off between the positive effect of node replacement at the initial stages of the outbreak (i.e., $\left[SI\right]$ to $\left[SS\right]$) and the negative impact it has once the disease is prevalent (i.e., $\left[II\right]$ to $\left[SI\right]$).
\subsection*{Analytical solution of the network model}
%
The evolution of the dynamical variables in the model is constrained by the conservation of nodes and links, namely
\begin{subequations} \label{eq:conservation_conditions}
\begin{align}
  [S] + [I] & = 1 \\
  \left[SS\right] + \left[SI\right] & + \left[II\right] = \frac{\left\langle k\right\rangle}{2} \ , \label{eq:links_conservation_condition}
\end{align}
\end{subequations}
where $\langle k \rangle$ is the average degree of nodes in the network. These conditions imply that
\begin{subequations} \label{eq:dot_conservation_conditions}
\begin{align}
  \dot{[S]} + \dot{[I]} & = 0 \\
  \dot{\left[SS\right]} + \dot{\left[SI\right]} & + \dot{\left[II\right]} = 0 \ ,
\end{align}
\end{subequations}
which allows us to effectively reduce Eqs.~\eqref{eq:unreduced_system} to the following system of three equations
\begin{subequations} \label{eq:reduced_system}
\begin{align}
  \dot{[I]} & = \beta \left[ SI \right] - r [I] \\
  \dot{\left[SS\right]} & = \left(r+\gamma\right) \left[SI\right] - 2\beta \left[SI\right] \frac{\left[SS\right]}{[S]} \\
  \dot{\left[II\right]} & = \beta \left[SI\right] \left( 1 + \frac{\left[SI\right]}{[S]} \right) - 2 \left(r+\gamma\right) \left[II\right] \ .
\end{align}
\end{subequations}
In other words, Eqs.~\eqref{eq:Sdot} and \eqref{eq:SIdot} can be dropped since the conservation conditions, Eqs.~\eqref{eq:conservation_conditions}, constrain the dynamics of variables $[S]$ and $[SI]$.
%
%
%
\subsubsection*{Steady state}
%

We can readily see that a \textit{disease-free} state, in which every node is susceptible,
\begin{align}
  [S]^*_\mathrm{df} = 1 \ ; \qquad [SS]^*_\mathrm{df} = \frac{\langle k \rangle}{2} \ ; \qquad [I]^*_\mathrm{df} = [SI]^*_\mathrm{df} = [II]^*_\mathrm{df} = 0 \ ,
\end{align}
is a steady state of the pair-wise network model [Eqs.~\eqref{eq:conservation_conditions} and Eqs.~\eqref{eq:reduced_system}]. To see whether there exists another steady state, i.e., an \textit{endemic} state in which a non-zero fraction of individuals are infectious, we set $\dot{[I]} = \dot{[SS]} = \dot{[II]} = 0$ in Eqs.~\eqref{eq:reduced_system}. This yields
\begin{subequations} \label{eq:steady_state_edge_density}
\begin{align}
  [SI]^*_\pm & = \frac{[I]^*_\pm}{a} \\
  [SS]^*_\pm & = \frac{1+b}{2a} (1-[I]^*_\pm) \\
  [II]^*_\pm & = \frac{[I]^*_\pm}{2(1+b)} \left( 1 + \frac{[I]^*_\pm}{a(1-[I]^*_\pm)} \right) \ ,
\end{align}
\end{subequations}
where we have defined the dimensionless parameters $a \equiv \beta/r$ and $b \equiv \gamma/r$. Substituting these last equations in Eq.~\eqref{eq:links_conservation_condition}, we obtain
\begin{align}\label{eq:steady_state_quadratic_equation}
  \frac{1+b}{2a} (1-[I]^*_\pm) + \frac{[I]^*_\pm}{a} + \frac{[I]^*_\pm}{2(1+b)} \left( 1 + \frac{[I]^*_\pm}{a(1-[I]^*_\pm)} \right) = \frac{\left\langle k\right\rangle}{2} \ ,
\end{align}
%
%
%
whose solutions,
\begin{align} \label{eq:steady_state}
  [I]^*_\pm = \frac{-\big[ (b+1)(a\langle k \rangle-2b) + a \big] \pm \sqrt{a^2(b\langle k \rangle+\langle k \rangle-1)^2-4a(b+1)(b\langle k \rangle-1)} }{2(b^2-a)}\ ,
\end{align}
correspond to the possible values of the endemic state. Note that only $[I]_-^*$ diverges when $a=b^2$ since $[I]_+^*$ converges towards
\begin{align}
  [I]_+^* = 1 - \frac{1}{b(b+1)(b\langle k \rangle - 1) - b} \ .
\end{align}
%
%
%
%
\subsubsection*{Stability analysis and bifurcation}
%

%
To determine whether the steady state solutions are stable or not, we linearize Eqs.~\eqref{eq:reduced_system} which yields the Jacobian matrix 
\begin{align}
  \mathbf{J} = r
  \begin{pmatrix}
    -1                                           &  -a                                                    &  -a \\
    \displaystyle -\frac{2a[SS][SI]}{(1-[I])^2}  &  \displaystyle -(1+b) - \frac{2a([SI] - [SS])}{1-[I]}  & \displaystyle -(1+b) + \frac{2a[SS]}{1-[I]} \\
    \displaystyle \frac{a[SI]^2}{(1-[I])^2}      &  \displaystyle -a - \frac{2a[SI]}{1-[I]}               & \displaystyle -2(1+b) - a - \frac{2a[SI]}{1-[I]}
  \end{pmatrix} \equiv r \mathbf{J^\prime}\ .
\end{align}
Notice that we have factorized $r$ from the matrix $\mathbf{J}$ to be able to work with the matrix $\mathbf{J^\prime}$ expressed solely in terms of the dimensionless parameters $a$ and $b$. Evaluating the matrix $\mathbf{J^\prime}$ at a given steady state value $[I]^*$ [using Eqs.~\eqref{eq:steady_state_edge_density}] and calculating the real part of its largest eigenvalue $\lambda_\mathrm{max}$, we conclude that the steady state is stable if $\lambda_\mathrm{max}<0$ and unstable if $\lambda_\mathrm{max}>0$.

The disease-free steady state undergoes a transcritical bifurcation whenever either $[I]^*_+=[I]^*_\mathrm{df}=0$ or $[I]^*_+=[I]^*_\mathrm{df}=0$, which happens when the constant term in Eq.~\eqref{eq:steady_state_quadratic_equation} equals to zero
\begin{align}
  (b+1)(b+1-a\langle k \rangle) = 0 \ .
\end{align}
Fixing either $a$ or $b$, we find the following threshold values
\begin{align} \label{eq:transcritical_thresholds}
  a^\mathrm{tr} = \frac{b+1}{\langle k \rangle} \ ; \qquad b^\mathrm{tr} = a\langle k \rangle - 1 \ .
\end{align}
Similarly, we see that the endemic steady state appears through a saddle-node bifurcation that occurs when $[I]_+^*=[I]_-^*$ which, from Eq.~\eqref{eq:steady_state}, happens when
\begin{align}
  a^2(b\langle k \rangle + \langle k \rangle - 1)^2 - 4a(b+1)(b\langle k \rangle - 1 ) = 0 \ .
\end{align}
Fixing either $a$ or $b$, we find the following threshold values

\begin{align} \label{eq:saddle-node_thresholds}
  a^\mathrm{sn} = \frac{4(b+1)(b\langle k \rangle - 1 )}{(b\langle k \rangle +\langle k \rangle - 1)^2} \ ; \qquad
  b_\pm^\mathrm{sn} = \frac{-(\langle k \rangle-1)(a\langle k \rangle-2)\pm2\sqrt{(\langle k \rangle+1)^2-a\langle k \rangle^2}}{\langle k \rangle(a\langle k \rangle-4)} \ .
\end{align}
Notice that whenever $a\langle k \rangle =4$, the threshold $b_-^\mathrm{sn}$ diverges but $b_+^\mathrm{sn}$ equals $-1-1/c(c-1)$ which is always negative and therefore can be discarded. From Eqs.~\eqref{eq:transcritical_thresholds} and \eqref{eq:saddle-node_thresholds}, we see that a bistable region appears or disappears when $a^\mathrm{tr}=a^\mathrm{sn}$ or $b^\mathrm{tr}=b^\mathrm{sn}_\pm$, which respectively yield the criteria
\begin{align} \label{eq:bistability_criteria}
  b > \frac{\langle k \rangle + 1}{\langle k \rangle} \ ; \qquad a > \frac{2\langle k \rangle + 1}{\langle k \rangle^2}
\end{align}
for the existence of a bistable region. 
%
%
%
\subsection*{Empirical analysis}
For exponential growth, the corresponding equation for the number of infected individuals as a function of elapsed time is:
\begin{align*}
P(t) = P_0(1+r)^t
\end{align*}
We estimated the parameters of both models from data using a non-linear least squares algorithm coded in the R programming language~\citep{R}. For both dengue and influenza we used data from the first eight weeks of each season.  We further explored the sensitivity of our results to the size of the training data and found them to be robust.  

\bibliography{teleportation}

\begin{thebibliography}{14}
\expandafter\ifx\csname natexlab\endcsname\relax\def\natexlab#1{#1}\fi
\expandafter\ifx\csname url\endcsname\relax
  \def\url#1{\texttt{#1}}\fi
\expandafter\ifx\csname urlprefix\endcsname\relax\def\urlprefix{URL }\fi

\bibitem[{Volz \emph{et~al.}(2011)Volz, Miller, Galvani \&
  Meyers}]{volz2011effects}
Volz, E.~M., Miller, J.~C., Galvani, A. \& Meyers, L.~A.
\newblock Effects of heterogeneous and clustered contact patterns on infectious
  disease dynamics.
\newblock \emph{PLoS Comput Biol} \textbf{7}, e1002042 (2011).

\bibitem[{Scarpino \emph{et~al.}(2015)}]{scarpino2015epidemiological}
Scarpino, S.~V. \emph{et~al.}
\newblock Epidemiological and viral genomic sequence analysis of the 2014 Ebola
  outbreak reveals clustered transmission.
\newblock \emph{Clinical Infectious Diseases} \textbf{60}, 1079--1082 (2015).

\bibitem[{Salath{\'e} \& Jones(2010)}]{salathe2010dynamics}
Salath{\'e}, M. \& Jones, J.~H.
\newblock Dynamics and control of diseases in networks with community
  structure.
\newblock \emph{PLoS Comput Biol} \textbf{6}, e1000736 (2010).

\bibitem[{Glass \emph{et~al.}(2006)Glass, Glass, Beyeler, Min
  \emph{et~al.}}]{glass2006targeted}
Glass, R.~J., Glass, L.~M., Beyeler, W.~E., Min, H.~J. \emph{et~al.}
\newblock Targeted social distancing design for pandemic influenza.
\newblock \emph{Emerg Infect Dis} \textbf{12}, 1671--1681 (2006).

\bibitem[{Fewtrell \emph{et~al.}(2005)}]{fewtrell2005water}
Fewtrell, L. \emph{et~al.}
\newblock Water, sanitation, and hygiene interventions to reduce diarrhoea in
  less developed countries: a systematic review and meta-analysis.
\newblock \emph{The Lancet infectious diseases} \textbf{5}, 42--52 (2005).

\bibitem[{Colón-González \emph{et~al.}(2011)Colón-González, Lake \&
  Bentham}]{colon}
Colón-González, F.~J., Lake, I.~R. \& Bentham, G.
\newblock Climate Variability and Dengue Fever in Warm and Humid {M}exico.
\newblock \emph{Am. J. Trop. Med. Hyg.} \textbf{84}, 757--763 (2011).

\bibitem[{Shaman \emph{et~al.}(2013)Shaman, Karspeck, Yang, Tamerius \&
  Lipsitch}]{shaman}
Shaman, J., Karspeck, A., Yang, W., Tamerius, J. \& Lipsitch, M.
\newblock Real-time influenza forecasts during the 2012–2013 season.
\newblock \emph{Nature Communications} \textbf{4} (2013).

\bibitem[{Shaman \emph{et~al.}(2010)Shaman, Pitzer, Viboud, Grenfell \&
  Lipsitch}]{shaman2010absolute}
Shaman, J., Pitzer, V.~E., Viboud, C., Grenfell, B.~T. \& Lipsitch, M.
\newblock Absolute humidity and the seasonal onset of influenza in the
  continental United States.
\newblock \emph{PLoS Biol} \textbf{8}, e1000316 (2010).

\bibitem[{Johansson \emph{et~al.}(2009)Johansson, Dominici, Glass
  \emph{et~al.}}]{johansson2009local}
Johansson, M.~A., Dominici, F., Glass, G.~E. \emph{et~al.}
\newblock Local and global effects of climate on dengue transmission in Puerto
  Rico.
\newblock \emph{PLoS Negl Trop Dis} \textbf{3}, e382 (2009).

\bibitem[{Wearing \& Rohani(2006)}]{wearing2006ecological}
Wearing, H.~J. \& Rohani, P.
\newblock Ecological and immunological determinants of dengue epidemics.
\newblock \emph{Proceedings of the National Academy of Sciences} \textbf{103},
  11802--11807 (2006).

\bibitem[{Watts \& Strogatz(1998)}]{watts1998collective}
Watts, D.~J. \& Strogatz, S.~H.
\newblock Collective dynamics of ‘small-world’networks.
\newblock \emph{nature} \textbf{393}, 440--442 (1998).

\bibitem[{Leventhal \emph{et~al.}(2015)Leventhal, Hill, Nowak \&
  Bonhoeffer}]{leventhal2015evolution}
Leventhal, G.~E., Hill, A.~L., Nowak, M.~A. \& Bonhoeffer, S.
\newblock Evolution and emergence of infectious diseases in theoretical and
  real-world networks.
\newblock \emph{Nature communications} \textbf{6} (2015).

\bibitem[{Gross \emph{et~al.}(2006)Gross, D’Lima \&
  Blasius}]{gross2006epidemic}
Gross, T., D’Lima, C. J.~D. \& Blasius, B.
\newblock Epidemic dynamics on an adaptive network.
\newblock \emph{Physical review letters} \textbf{96}, 208701 (2006).

\bibitem[{{R Core Team}(2015)}]{R}
{R Core Team}.
\newblock \emph{R: A Language and Environment for Statistical Computing}.
\newblock R Foundation for Statistical Computing, Vienna, Austria (2015).
\newblock \urlprefix\url{http://www.R-project.org/}.

\end{thebibliography}

\section*{Acknowledgements}
S.V.S. received funding support from the Santa Fe Institute and the Omidyar Group.  A.A. received funding support from the Fonds de recherche du Qu\'ebec -- Nature et technologies and the James S. McDonnell Foundation. L.H.-D. received funding support from the James S. McDonnell Foundation Postdoctoral Fellowship and the Santa Fe Institute.  

\section*{Author contributions}
S.V.S and L.H.D. conceived the project; L.H.D and A.A. performed the simulations and calculations; S.V.S. analyzed the empirical data; all authors interpreted the results and produced the final manuscript.

\section*{Conflicts of interest}
The authors declare no conflicts of interest exist.

\end{document}